# Predicting the Inorganic Exciton Peak Position in 2D Hybrid Organic-Inorganic Perovskites from Hybrid Density Functional Theory

Svenja M. Janke[1]

[1]Department of Chemistry, University of Warwick, Coventry, United Kingdom

## TOC

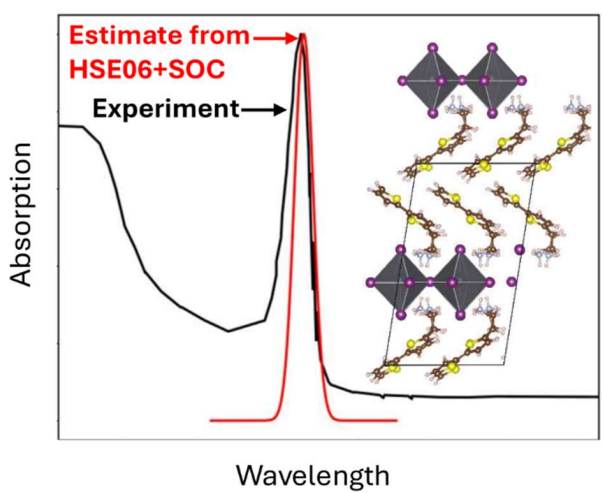

## Abstract

Modelling the inorganic exciton contribution to 2D hybrid organic-inorganic perovskites is essential to understand their properties and screen for new materials. Here, we combine hybrid density functional theory calculations including spin orbit coupling (SOC) with the experimental relationship between the inorganic band gap and exciton binding energy to predict the inorganic exciton energy. For this purpose, we determine a universal exchange parameter for the HSE06 hybrid functional with SOC for lead-based 2D hybrid organic-inorganic perovskites. We further identify a relationship that connects PBE calculations to experiment-quality optical gaps and allows us to generalize the exchange mixing parameter other SOC approximations. Our approach opens the path to screen for 2D hybrid organic-inorganic perovskites with optimized spectra, e.g., for new solar cell or light emitting materials.

## Main Text

Hybrid organic-inorganic perovskites (HOIP) have in recent years attracted great interest as potential solar cell materials associated with their rapid rise in light harvesting performance[1].  For 3D HOIPs, only a limited compositional space is available. Replacing the small organic cation with optically active organic molecules results in lower-dimensional perovskites like 2D HOIPs and opens the path to further tune and enhance the material's functionality: The organic molecules can stabilize the HOIP[2] and could assist light harvesting or light emission[3,4]. To understand design principles or screen for new, more energy-efficient solar cell or light emitting device (LED) materials, it is essential to model the materials' absorption and emission spectra.

The dominating contributions to 2D HOIP spectra in the visible range arise from excitons on the inorganic and, for 2D HOIPs with optically active organic molecules, excitons on the organic component. In principle, excitonic properties are accessible from Bethe-Salpeter equation (BSE) or linear-response time-dependent (TD) density functional theory (DFT)[5]. But in practice, 2D HOIPs with optically active cations have supercells with hundreds of atoms. Additionally, the exciton on the organic component couples to the lattice vibrations of the organic molecules, leading to distinctive vibronic progressions in the absorption spectra. Without further approximations, these large supercells and the necessity to include lattice vibrations make using BSE and TD-DFT to model spectra prohibitively computationally expensive. As a first step to solving this problem, we recently showed that trends in the vibronic progression in the organic contribution to 2D HOIP absorption spectra can be related to trends in the structure of the organic component using a Frenkel-Holstein Hamiltonian (FHH)[6].

But until now, predicting the inorganic exciton energetic position in large 2D HOIP absorption and emission spectra from first principles theory has remained out of computational reach. In this work, we combine first principles theory at the level of hybrid DFT including spin-orbit coupling (SOC) with an established experimental relation[7] and derive a general exchange mixing parameter for lead-based 2D HOIPs to predict the inorganic exciton contribution to 2D HOIP spectra.

Hansen et al.[7] recently reported a strong linear relationship between the measured inorganic exciton binding energy $E_b$ and the band gap $E_g$ in 2D HOIPs. This experimental relation deviates from the linear scaling relation of $E_b \approx 0.25 E_g$ for 2D semiconductors derived from BSE and ***k.p*** theory calculations[8] due to partial screening by phonons[7]:

$$E_b = 0.18 E_g - 0.25 \text{ eV} \tag{1}$$

In principle, the experimental band gap $E_g$ could be approximated using the Kohn-Sham band gap from hybrid DFT+SOC calculations $E_g^{KS}$. The HSE06 hybrid DFT functional[9] combined with 2nd variational non-self-consistent SOC[10] is an excellent candidate for this purpose. HSE06+SOC has been highly successful at explaining where charge carriers are located after light absorption or whether a material can emit light from the organic or inorganic component based on qualitatively modelling level alignment of the organic and inorganic frontier levels, see e.g., Refs.[11–18]. To investigate if $E_g^{KS}$ from HSE06+SOC can be used to predict the inorganic exciton position using Eq. (1), we have assembled a set of 16 2D HOIPs $(A)_n PbX_4$ (n = 1 or 2; X = Cl, Br, I; for a full list see Figure S1, Table S1) based on three criteria. A) the availability of experimental data on the absorption or emission inorganic exciton peak position (i.e., the inorganic exciton energy), B) the availability of atomic structures that have previously been found to reproduce experimental observations well (lattice parameters or level alignment), or atomic structures without disorder for the organic cation, C) a range of structural variability in the organic cation and halide anions (see Figure S1). We include lead-iodine-, lead-bromine- and lead-chlorine-based perovskites and organic cations forming Dion-Jacobsen (n = 1) and Ruddlesden-Popper (n = 2) 2D HOIPs, with methyl- and ethyl- (i.e., short and long) linker chains to the inorganic component, aromatic cations with and without heteroatoms and aliphatic organic cations. Atomic structures were already available for (AE4T)PbCl$_4$[11,19], (PMA)$_2$PbCl$_4$[20], (NMA)$_2$PbCl$_4$[20], (R-NEA)$_2$PbBr$_4$[15], (PMA)$_2$PbBr$_4$[20], (NMA)$_2$PbBr$_4$[20], (AE4T)PbBr$_4$[11,19], (AE4T)PbI$_4$[11,19], (AE2T)PbI$_4$[12], (2T)$_2$PbI$_4$[13], (BTm)$_2$PbI$_4$[13], (PEA)$_2$PbI$_4$[17], (PMA)$_2$PbI$_4$[20]. For (PEA)$_2$PbBr$_4$[21], room-temperature (BA)$_2$PbI$_4$[22] and (EOA)$_2$PbI$_4$[23], where experimental structures without disorder are available, we perform geometry optimizations using the GGA Perdew–Burke–Ernzerhof (PBE) functional[24] with Tkatchenko-Scheffler van-der-Waals correction[25] (TS) within the FHI-aims all electron electronic structure code[10,26–30]. PBE+TS has been shown to give reliable approximations for the atomic structure in comparison to experiment[11]. We relax the initial structure until forces are converged up to 0.005 eV/A with FHI-aims's 'tight' settings. We find (Table S2) that the relaxed lattice parameters are in good agreement and within 2% of the experimental lattice parameter and angles for all but (BA)$_2$PbI$_4$, where the lattice parameter in c-direction disagrees by 2.4 %. Our results are in line with similar uncertainties of previous 2D HOIP structure optimizations[11,15,17]. We have previously[6] shown that the contribution of an organic Frenkel exciton to 2D HOIP absorption spectra can be modeled with a Frenkel-Holstein Hamiltonian. Given the significant effort that is currently still associated with setting up the Frenkel-Holstein Hamiltonian for a material and the focus of the current work on the inorganic exciton, we will not include modelling of the organic exciton contribution here.

For all 16 2D HOIPs, we calculated the Kohn-Sham band gap $E_g^{KS}$ with HSE06, 2$^{nd}$ variational non-self-consistent SOC and 'intermediate' settings (Table S3). To compensate for the delocalization error and the underestimation of experimental band gaps, the HSE06 functional mixes short-range Hartree Fock exchange $E_x^{HF,SR}$ with short-range exchange from the PBE functional $E_x^{PBE,SR}$ [9]:

$$E_{xc}^{HSE06} = \beta E_x^{HF,SR}(\omega) + (1-\beta)E_x^{PBE,SR}(\omega) + E_x^{PBE,LR}(\omega) + E_C^{PBE} \qquad (2)$$

$E_x^{PBE,LR}$ is the short- and long-range components of the PBE exchange functional and $E_C^{PBE}$ is the PBE correlation energy. The screening parameter $\omega$ = 0.11 bohr$^{-1}$ defines the separation range. The exchange mixing parameter $\beta$ controls the mixing ratio Hartree Fock to PBE exchange, with $\beta_{default} = 0.25$. $\beta$ is material specific and increasing $\beta$ increases the Kohn-Sham band gap[31,32].

We use Eq. (1) to estimate the band gap $E_g^{est}$ from experimental inorganic exciton peak positions from absorption or emission spectra (Table S1 and S3) and then calculated the Kohn Sham band gap $E_g^{KS}$ with HSE06's default setting and SOC (for more details on the computational methods, see supplementary material). Comparing $E_g^{KS}$ to $E_g^{est}$, we find that $E_g^{KS}$ underestimates $E_g^{est}$ on average by 640 meV. $E_g^{KS}$ from HSE06+SOC with its default settings hence cannot be used to estimate the inorganic exciton peak position.

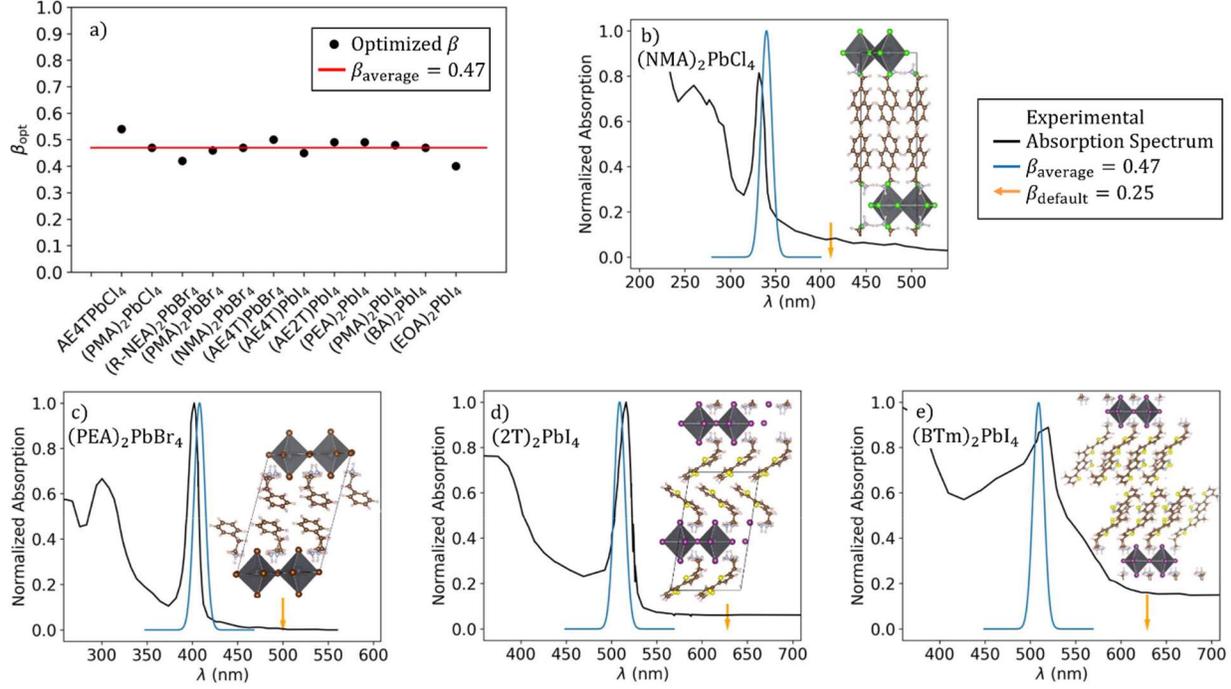

**Figure 1**: a) The optimized $\beta_{\text{opt}}$ for the twelve sampled perovskites; b-e) experimental absorption spectra[7,13,33] (black), estimate of inorganic exciton position obtained with Eq. (1) and the band gap of $\beta_{\text{average}}$ (blue, Gaussian peak with 6 nm broadening) and $\beta_{\text{default}} = 0.25$ (orange arrow) for the 2D HOIPs in the test set b) $(NMA)_2PbCl_4$, c) $(PEA)_2PbBr_4$, d) $(2T)_2PbI_4$ and e) $(BTm)_2PbI_4$. The insets show the 2D HOIPs unit cells visualized with VESTA[34].

We next investigate if there is a universal $\beta$ to model the inorganic band gap of 2D HOIPs. For this purpose, we split the dataset of the 16 HOIP into a training set of twelve and a test set of four 2D HOIPs (one for X = Cl, one for X = Br and 2 for X = I, randomly picked from the subsets). For the training set, we vary $\beta$ until the match between the HSE06+SOC $E_g^{\text{KS}}$ and $E_g^{\text{est}}$ cannot be improved within $\Delta\beta = \pm 0.01$. Figure 1a shows that the optimized $\beta_{\text{opt}}$ for the training set lies between $\beta = 0.40$ and 0.54. The average of the optimized $\beta$ is $\beta_{\text{average}} = 0.47$ with a maximum deviation of $\Delta\beta = 0.07$. $\Delta\beta = 0.07$ corresponds to a maximum deviation of the band gap of 190 meV, a sizable improvement over the average deviation of 640 meV when using HSE06+SOC with its default $\beta_{\text{default}} = 0.25$. However, 190 meV still means a deviation of ~21 nm on the high energy (750 nm) and ~99 nm on the low-energy end (380 nm) of visual spectrum. To estimate how large the variance of $\beta_{\text{average}}$ is, that is, to estimate how much $\beta$ can be expected to deviate from 0.47 when applying it to a new 2D HOIP, we perform a bootstrap. We resample the training set 10000 times, which is enough to converge the standard deviation $\sigma = 0.01$ and a variance of $\text{Var}(\beta_{\text{average}}) = 0.0001$. Figure S2 shows the histogram resulting from the bootstrap. Based on this bootstrap, the optimal $\beta$ for other 2D HOIPs is likely to be close to $\beta_{\text{average}} = 0.47$.

Figure 1b-e shows the inorganic exciton peak calculated using HSE06+SOC with $\beta_{\text{average}}$ and Eq. (1) and applying a Gaussian line shape function with a broadening of 6 nm for the test set 2D HOIPs (blue) in comparison to the experimental spectra[7,13,33] (black). The position of the inorganic exciton differs by 5 nm, 5 nm, 7 nm and 8 nm for $(NMA)_2PbCl_4$, $(PEA)_2PbBr_4$, $(2T)_2PbI_4$ and $(BTm)_2PbI_4$, respectively, a minimal color change and a significant improvement over the absorption spectrum estimated with $\beta_{\text{default}} = 0.25$, the default HSE06 value (orange arrow in Fig. 1b-e). Using Eq. (1) with HSE06 and $\beta_{\text{average}} = 0.47$ and 2nd variational non-self-consistent SOC in FHI-aims gives hence a reasonable approximation of the position of the inorganic exciton in the optoelectronic spectrum. The universality of $\beta_{\text{average}}$ over our data set is likely a result of the fact that the inorganic frontier orbitals of group 14-based HOIPs have the same composition. Halide-*p* and metal-*s* orbitals make up the valence band and halide-*p* and Pb-*p* the conduction band[35,36]. Finally, in Figure S3 in the supplementary material, we additionally show that the optimized $\beta$ retains HSE06+SOC's ability to qualitatively explain experimental level alignment.

To obtain an even better approximation for $\beta$ for 2 D HOIPs or beyond the SOC approximation used in this work, it would be advantageous if $\beta$ could be estimated based on PBE calculations. The high-field dielectric constant $\varepsilon_\infty$ of semiconducting materials have been observed to correlate with the inverse Hartree Fock mixing parameter, see e.g., Ref.[31]. For 2D HOIPs, $\varepsilon_\infty$ of the inorganic component has been estimated to be $\varepsilon_{\infty,\text{inorg}} = 2.87$ for Cl-based 2D HOIPs, $\varepsilon_{\infty,\text{inorg}} = 3.55$ for Br-based 2D HOIPs and $\varepsilon_{\infty,\text{inorg}} = 4.53$ for I-based 2D HOIPs[7]. Given that the different halide anions Cl, Br and I do not show significant differences in $\beta_{\text{opt}}$, there is clearly no correlation between $1/\beta_{\text{opt}}$ and $\varepsilon_{\infty,\text{inorg}}$. But one reason for $\beta$ being slightly different for different 2D HOIPs could be the different dielectric environments caused by changing the organic cation. For example, for the perovskites with the (AE4T) cation, $\beta_{\text{opt}} = 0.54$, 0.50 and 0.45 for the halide anions Cl, Br, and I, respectively, which correlates with the decreasing packing density of the (AE4T) cation[6]. To investigate if a correlation with $1/\beta_{\text{opt}}$ can be found, we obtain $\varepsilon_\infty$ for the entire 2D HOIP system ($\varepsilon_{\infty,\text{total}}$), for the organic ($\varepsilon_{\infty,\text{org}}$) and the inorganic component ($\varepsilon_{\infty,\text{inorg}}$) from the dielectric profile using the procedure outlined in Ref.[37]. However, we also do not observe any correlation between $\beta_{\text{opt}}$ and $\varepsilon_{\infty,\text{total}}$ or $\beta_{\text{opt}}$ and the dielectric mismatch $\varepsilon_{\infty,\text{org}}/\varepsilon_{\infty,\text{inorg}}$ (see Figure S4). In addition, we observe a very small dielectric mismatch for the 16 2D HOIPs in our study, with close to identical $\varepsilon_{\infty,\text{org}}$ and $\varepsilon_{\infty,\text{inorg}}$ for $(PMA)_2PbCl_4$, $(NMA)_2PbCl_4$, $(R\text{-}NEA)_2PbBr_4$, $(NMA)_2PbBr_4$ and $(AE4T)PbI_4$ (see Table S4 and S5). Our findings thus further support Hansen's[38] recent observation that the dielectric mismatch between organic and inorganic component in 2D HOIPs is small and has little correlation with inorganic exciton binding energy in 2D HOIPs. Additionally, our investigations make clear that a correlation between $1/\beta_{\text{opt}}$ and $\varepsilon_\infty$ does not apply to 2D HOIPs.

Figure 2a shows that for all 2D HOIP in the training set, the Kohn-Sham band gap $E_g^{\text{KS}}$ depends linearly on $\beta$ with very similar slopes $m_\beta$. From Eq. (2), it further follows that for $\beta$ = 0.0, the energy will only contain (screened) PBE contributions. The band gap at $\beta = 0.0$ can hence be approximated to correspond to the PBE+SOC band gap. Accordingly, $\beta_{\text{opt}}$ could be approximated ($\beta_{\text{lin}}$) from the estimated experimental band gap $E_g^{\text{est}}$ and the energy of the PBE+SOC band gap $E_g^{\text{PBE+SOC}}$ as

$$E_g^{\text{est}} = m_\beta \beta_{\text{lin}} + E_g^{\text{PBE+SOC}} . \qquad (3)$$

If $m_\beta$ and $E_g^{\text{est}}$ are known, any SOC approximation could be used to calculate $E_g^{\text{PBE+SOC}}$ and to obtain an approximation for $\beta$ in HSE06, making it possible to go beyond the 2nd variational non-self-consistent SOC used in this work.

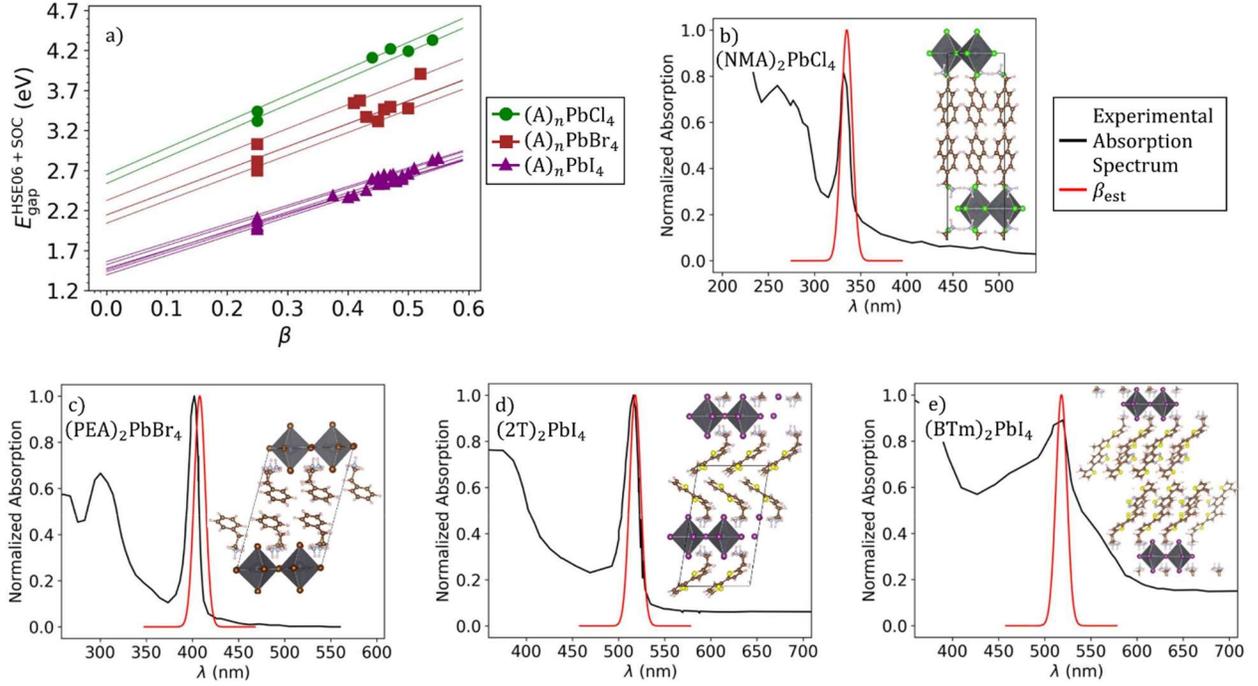

**Figure 2**: a) HSE06+SOC band gap as a function of $\beta$ for the training set for the Cl-based 2D HOIP (green points), the Br-based 2D HOIPs (brown points) and the I-based 2D HOIPs (purple points). The 2D HOIPs have similar slopes for each of the halide anions Cl (green lines), Br (brown lines) and I (purple lines). b)-e) The inorganic exciton obtained from Eq. (1) and estimating $\beta_{lin}$ from the PBE+SOC band gap (red) and $E_g^{est}$, compared to the experimental absorption spectra[7,13,33] (black) for the test set 2D HOIPs b) $(NMA)_2PbCl_4$, c) $(PEA)_2PbBr_4$, d) $(2T)_2PbI_4$ and e) $(BTm)_2PbI_4$. The insets show the 2D HOIPs unit cells.

**Table 1**: Slope $m_\beta$ for different halide anions to estimate $\beta$ ($\beta_{lin}$) with Eq. (3) using the estimated experimental band gap $E_g^{est}$ and the PBE+SOC band gap energy from the training set. The last column shows the root-mean-square error between $\beta_{lin}$ and $\beta_{opt}$.

| Halide | $m_\beta$ (eV) | RMSE ($\beta$) |
|---|---|---|
| Cl | 3.30 | 0.00 |
| Br | 2.88 | 0.00 |
| I  | 2.37 | 0.01 |

**Table 2**: For the test set perovskites, we use Eq. (3) with the slopes $m_\beta$ from Table 1 to obtain $\beta_{lin}$ from the PBE+SOC band gap. The last column shows the optimal $\beta_{opt}$ obtained by tuning $\beta$ to reproduce $E_g^{est}$.

| Perovskite | $\beta_{lin}$ | $\beta_{opt}$ |
|---|---|---|
| $(NMA)_2PbCl_4$ | 0.49 | 0.49 |
| $(PEA)_2PbBr_4$ | 0.47 | 0.475 |
| $(2T)_2PbI_4$ | 0.45 | 0.46 |
| $(BTm)_2PbI_4$ | 0.45 | 0.45 |

To obtain $m_\beta$, we perform linear regressions for the 2D HOIPs in the training set and average over the slope for each halide anion. Table 1 shows the slope $m_\beta$ we obtain for Eq. (3) for the different halide anions in the training set. Using Eq. (3) with the $m_\beta$ from Table 1 to approximate $\beta_{\text{opt}}$ leads to very low root-mean-square errors for the training set. Table 2 shows that in applying Eq. (3) to the test set, the estimated $\beta_{\text{lin}}$ recovers the optimal $\beta_{\text{opt}}$ to reproduce the experimental band gap $E_g^{\text{est}}$ within $\Delta\beta \pm 0.01$. As a result, Figure 2a-d shows that HSE06+SOC with $\beta_{\text{lin}}$ and using Eq. (1) gives a very good approximation for the inorganic exciton peak position.

In summary, we predict the inorganic exciton peak position in $Pb^{2+}$-based 2D HOIP spectra by combining the established experimental relationship between 2D HOIP exciton binding energy and the band gap, and first principles theory. For this purpose, we determine a 2D HOIP-universal exchange mixing parameter for HSE06+SOC of $\beta_{\text{average}} = 0.47$. We further identify a linear relationship in Eq. (3) that connects PBE+SOC calculations to the experiment-derived band gap and permits determining the exchange mixing parameter for other spin-orbit coupling approximations. Because the universality of $\beta$ for $Pb^{2+}$-based 2D HOIPs is presumably a result of the analogous orbital composition of their inorganic frontier levels, we expect that Eq. (3) can also be used to estimate $\beta$ for systems with similar inorganic frontier level make up like[36] $Sn^{2+}$, $Ge^{2+}$, $Sb^{3+}$ and $Bi^{3+}$-based 2D HOIPs.

Our work enables predicting the inorganic exciton contribution to $Pb^{2+}$-based spectra of large 2D HOIPs, making it possible to understand experimental spectra in greater depth. The relatively small computational cost of our approach in comparison to BSE of TDDFT calculations further opens the pathway to screen for 2D HOIPs with optimized spectra, e.g., for photovoltaic or LED application. We will in future combine our approach with our precious work on modelling the organic exciton contribution to absorption spectra[6] to further understand and screen for new optical materials.

# Supplementary Materials

The supplementary material contains a figure of the organic cations and halide anions in test and training set of 16 2D HOIP; a table of the list of 16 2D HOIPs in test and training set; a table of the lattice parameters and angles from structure relaxation; a table of the inorganic band gaps for different $\beta$ for test and training set; a summary of computational methods; a figure of the bootstrap for the Optimized Hartree Fock Exchange Mixing Parameter β; a figure of the level Alignment for HSE06+SOC with β=0.25 and β$_{\text{opt}}$; a discussion of the estimate of high-field dielectric constant of organic and inorganic component and its correlation with β$_{\text{opt.}}$; a figure of the correlation of the total dielectric constant and dielectric mismatch with 1/ β; a table the high-field effective dielectric constants from dielectric profile for t$_1$ = 0 Å integration border selection; a table the high-field effective dielectric constants from dielectric profile for t$_1$ = 0.1 Å integration selection.

# Acknowledgement

SMJ gratefully acknowledges financial support from the Ramsay Memorial Fellowship Trust and the University of Warwick for support through a Ramsay Fellowship. Via our membership of the UK's HEC Materials Chemistry Consortium, which is funded by EPSRC (EP/X035859), this work used the ARCHER2 UK National Supercomputing Service (http://www.archer2.ac.uk). Calculations were performed using the

Sulis Tier 2 HPC platform hosted by the Scientific Computing Research Technology Platform at the University of Warwick. Sulis is funded by EPSRC Grant EP/T022108/1 and the HPC Midlands+ consortium. SMJ gratefully acknowledges University of Warwick's SCRTP high performance computing facilities, in particular Avon, and support. SMJ thanks Dr April Bialas, Dr Mohammed Sejour and Dr Phoebe Askeson, Prof. Reinhard Maurer, Dr Zsuzsanna Koczor-Benda, Dr Lukas Hörmann, Dr Nils Hertl and Prof. Volker Blum for useful discussions.

# Supplementary Material
# Predicting the Inorganic Exciton Peak Position in 2D Hybrid Organic-Inorganic Perovskites from Hybrid Density Functional Theory

Svenja M. Janke[1]

[1]Department of Chemistry, University of Warwick, Coventry, United Kingdom

Contents:

Table of Contents

# 1. Figure S1: Organic Cations and Halide Anions used in Test and Training Set

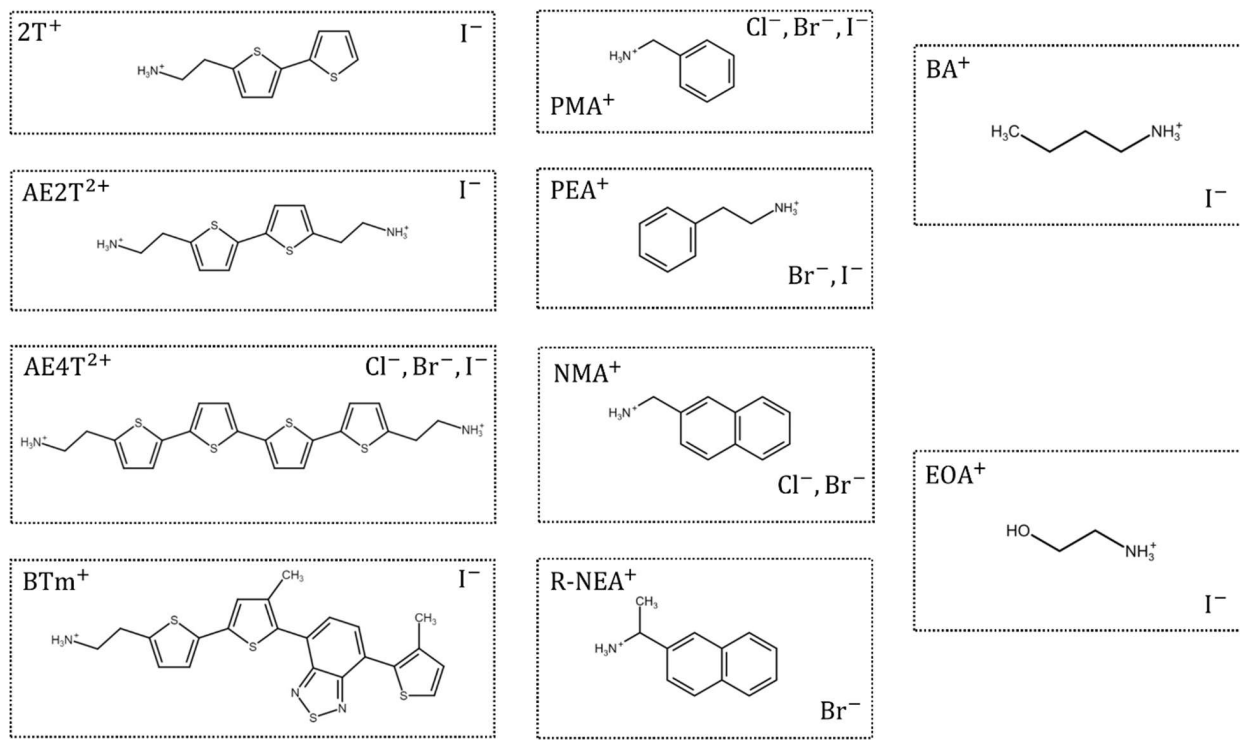

**Figure S1**: The halide anions of Cl, Br and I are paired in $Pb^{2+}$-based 2D HOIPs with the organic cations 2T: 2-([2,2 0 -bithiophene]-5-yl)ethylammonium, AE2T: 5,5'-diylbis(aminoethyl)-[2,2'-bithiophene], AE4T: 5,5'''-bis(aminoethyl)-2,2':5',2'':5'',2'''-quaterthiophene, BTm: tert-Butyl(2-(4'-methyl-5'-(7-(3-methylthiophen-2-yl)benzo[c][1,2,5]thiadiazol-4-yl)-[2,2'bithiophen]-5-yl)ethyl) PMA: phenylmethylammonium, PEA: 2-phenylethylammonium, NMA: 1-(2-naphthyl)-methanammonium, R-NEA: R-1-(1-naphthyl)ethylamine), BA: butylammonium, EOA: Ethanolamine. The perovskites in the data set are: 1. $(AE4T)PbCl_4$[11,19], 2. $(PMA)_2PbCl_4$[20], 3. $(NMA)_2PbCl_4$[20], 4. $(R-NEA)_2PbBr_4$[15], 5. $(PEA)_2PbBr_4$[21], 6. $(PMA)_2PbBr_4$[20], 7. $(NMA)_2PbBr_4$[20], 8. $(AE4T)PbBr_4$[11,19], 9. $(AE4T)PbI_4$[11,19], 10. $(AE2T)PbI_4$[12], 11. $(2T)_2PbI_4$[13], 12. $(BTm)_2PbI_4$[13], 13. $(PEA)_2PbI_4$[17], 14. $(PMA)_2PbI_4$[20], 15. Room-temperature $(BA)_2PbI_4$[22] and 16. $(EOA)_2PbI_4$[23].

## 2. Table S1: Overview of 2D HOIPs in Test and Training Set

**Table S1**: 2D HOIPs investigated for the test and training set with the references to the atomic structural model that were used in the benchmark. If 'relaxed' is 'n', the atomic structural model from the indicated reference was used, if 'y', the atomic structural model from the indicated reference was relaxed. The relaxed lattice parameters are given in Table S2. The energy of the inorganic excitonic peak for absorption and emission is indicated. We have used the absorption energy for our work wherever available to determine the extrapolated band gap using Eq. (1) in the main text to benchmark HSE06+SOC band gaps.

| | Perovskite | Relaxed? | Reference atomic structural model | Energy of excitonic Peak ($^a$=absorption, $^e$=emission) nm (eV) | Reference Excitonic peak energy | Band gap (eV) $^{extra}$=extrapolated $E_g^{est}$ using Eq. 1 in main text (eV) |
|---|---|---|---|---|---|---|
| 1 | (AE4T)PbCl$_4$ | n | [11] | 327$^a$ (3.79) | [12,19] | 4.32$^{extra}$ |
| 2 | (PMA)$_2$PbCl$_4$ | n | [20] | 335.536$^a$ (3.70) 333.3$^e$ (3.72) | [33] | 4.20$^{extra}$ |
| 3 | (NMA)$_2$PbCl$_4$ | n | [20] | 335.314$^e$ (3.70) | [33] | 4.21$^{extra}$ |
| 4 | R-(NEA)$_2$PbBr$_4$ | n | [15] | 391 (3.17$^e$) | [15] | 3.56$^{extra}$ |
| 5 | (PEA)$_2$PbBr$_4$ | y | [21] | | | 3.39$^7$ |
| 6 | (PMA)$_2$PbBr$_4$ | n | [20] | 402.938$^e$ (3.08) | [33] | 3.45$^{extra}$ |
| 7 | (NMA)$_2$PbBr$_4$ | n | [20] | 392.625$^a$ (3.16) 398.6$^e$ (3.11) | [33] | 3.55$^{extra}$ |
| 8 | (AE4T)PbBr$_4$ | n | [11] | 403$^a$ (3.08) | [19] | 3.45$^{extra}$ |
| 9 | (AE4T)PbI$_4$ | n | [11] | 515$^a$ (2.41) | [12,19] | 2.63$^{extra}$ |
| 10 | (AE2T)PbI$_4$ | n | [12] | 518.5$^a$ (2.39) 526$^e$ (2.36) | [12] | 2.61$^{extra}$ |
| 11 | (2T)$_2$PbI$_4$ | n | [13] | 518.4$^e$ (2.39) 516.148$^a$ (2.40) | [13] | 2.63$^{extra}$ |
| 12 | (BTm)$_2$PbI$_4$ | n | [13] | 517$^a$ (2.40) | [13] | 2.62$^{extra}$ |
| 13 | (PEA)$_2$PbI$_4$ | n | [17] | 2.40$^e$ | [17] | 2.62$^{extra}$ |
| 14 | (PMA)$_2$PbI$_4$ | n | [20] | 525.743$^e$ (2.36) | [33] | 2.57$^{extra}$ |
| 15 | (BA)$_2$PbI$_4$ | y | [22] | | | 2.83$^7$ |
| 16 | (EOA)$_2$PbI$_4$ | y | [23] | | | 2.38$^7$ |

# 3. Table S2: Lattice Parameters and Angles from Structure Relaxation

**Table S2**: Lattice parameters and angles from the structure relaxation in comparison to the experimental results. Initial structures were taken from the indicated experimental references. The structures were relaxed with PBE+TS in FHI-aims[26], first with `light' settings and then with 'tight' settings until forces were converged below 0.005 eV/Å.

|  | a (A) | b (A) | c (A) | α (°) | β (°) | γ (°) | Ref |
|---|---|---|---|---|---|---|---|
| $(BA)_2PbI_4$ | | | | | | | |
| Exp. RT | 8.8763(3) | 8.6936(3) | 27.6162(13) | 90 | 90 | 90 | [22] |
| Theory | 8.682 | 8.581 | 28.302 | 90 | 90 | 90 | |
| Deviation (%) | 2.2 | 1.3 | 2.4 | 0 | 0 | 0 | |
| $(EOA)_2PbI_4$ | | | | | | | |
| Exp. | 8.935(1) | 9.056(2) | 10.214(3) | 90 | 100.26(1) | 90 | [23] |
| Theory | 8.946 | 8.996 | 10.209 | 90 | 100.2 | 90 | |
| Deviation(%) | 0.1 | 0.7 | 0.0 | 0.0 | 0.1 | 0.0 | |
| $(PEA)_2PbBr_4$ | | | | | | | |
| Exp. | 11.6150(4) | 11.6275 (5) | 17.5751 (6) | 99.5472 (12) | 105.7245 (10) | 89.9770 (12) | [21] |
| Theory | 11.555 | 11.431 | 17.391 | 99.4 | 106.4 | 90.0 | |
| | 0.5 | 1.7 | 1.0 | 0.2 | 0.6 | 0.0 | |

# 4. Table S3: Inorganic band gap for different $\beta$ for test and training set

**Table S3**: Inorganic band gap obtained for different $\beta$ for all 2D HOIPs. With $b = E_g^{\text{PBE+SO}}$ and the slope $m_\beta$ resulting from the linear regression for Eq (3) in the main text. $\beta_{\text{opt}}$ is bolded.

*Value from Ref.[11] **Value from Ref. [17].

| Perovskite | $\beta$ | $E_g^{\text{HSE06+SOC}}$ (intermediate) [eV] | $E_g^{\text{HSE06+SOC}}$ (tight) [eV] | $E_g^{\text{PBE+SOC}}$ [eV] | $m_\beta$ (eV) |
|---|---|---|---|---|---|
| One perovskite for several rows of beta | | | | | |
| (AE4T)PbCl₄ | 0.25 | 3.320* | | 2.537 | 3.29 |
| | 0.5 | 4.195 | 4.195 | | |
| | **0.54** | 4.332 | | | |
| (PMA)₂PbCl₄ | 0.25 | 3.439 | | 2.648 | 3.31 |
| | 0.44 | 4.113 | | | |
| | **0.47** | 4.222 | | | |
| (NMA)₂PbCl₄ | 0.25 | 3.372 | | 2.590 | Test set |
| | 0.46 | 4.096 | | | |
| | 0.47 | 4.131 | | | |
| | **0.49** | 4.201 | | | |
| | 0.50 | 4.235 | | | |
| R-(NEA)₂PbBr₄ | 0.25 | 3.028 | | 2.325 | 3.00 |
| | 0.41 | 3.541 | | | |
| | **0.42** | 3.574 | | | |
| | 0.52 | 3.907 | | | |
| (PEA)₂PbBr₄ | 0.25 | 2.717 | | 2.055 | Test set |
| | 0.375 | 3.097 | | | |
| | 0.45 | 3.332 | | | |
| | 0.47 | 3.395 | | | |
| | **0.475** | 3.412 | | | |
| | 0.48 | 3.427 | | | |
| (PMA)₂PbBr₄ | 0.25 | 2.813 | 2.813 | 2.143 | 2.86 |
| | 0.43 | 3.369 | 3.369 | | |
| | **0.46** | 3.464 | | | |
| | 0.47 | 3.496 | | | |

| | | | | | | |
|---|---|---|---|---|---|---|
| (NMA)₂PbBr₄ | 0.25 | 2.815 | 2.815 | | 2.144 | 2.85 |
| | **0.47** | 3.497 | | | | |
| (AE4T)PbBr₄ | 0.25 | 2.700* | | | 2.040 | 2.84 |
| | 0.45 | 3.316 | 3.316 | | | |
| | **0.50** | 3.475 | | | | |
| (AE4T)PbI₄ | 0.25 | 2.110* | | | 1.561 | 2.35 |
| | 0.44 | 2.596 | | | | |
| | **0.45** | 2.622 | | | | |
| | 0.46 | 2.649 | | | | |
| (AE2T)PbI₄ | 0.25 | 2.022 | | | 1.476 | 2.29 |
| | 0.45 | 2.531 | | | | |
| | 0.46 | 2.557 | | | | |
| | 0.48 | 2.566 | | | | |
| | **0.49** | 2.593 | | | | |
| (2T)₂PbI₄ | 0.25 | 2.101 | | | 1.551 | Test set |
| | 0.44 | 2.586 | | | | |
| | 0.45 | 2.612 | | | | |
| | **0.46** | 2.637 | | | | |
| | 0.47 | 2.665 | | | | |
| | 0.48 | 2.692 | | | | |
| (BTm)₂PbI₄ | 0.25 | 2.099 | | | 1.548 | Test set |
| | 0.44 | 2.584 | | | | |
| | **0.45** | 2.611 | | | | |
| | 0.47 | 2.664 | | | | |
| | 0.48 | 2.691 | | | | |
| (PEA)₂PbI₄ | 0.25 | 2.01** | | | 1.459 | 2.41 |
| | **0.49** | 2.632 | | | | |
| | 0.50 | 2.659 | | | | |
| | 0.51 | 2.728 | | | | |
| (PMA)₂PbI₄ | 0.25 | 1.990 | | | 1.435 | 2.38 |
| | 0.43 | 2.455 | | | | |
| | 0.46 | 2.535 | | | | |
| | **0.48** | 2.589 | | | | |
| | 0.49 | 2.616 | | | | |
| (BA)₂PbI₄ (RT) | 0.25 | 2.073 | | | 1.523 | 2.39 |
| | 0.375 | 2.389 | | | | |
| | **0.47** | 2.639 | | | | |
| | 0.475 | 2.652 | | | | |
| | 0.54 | 2.828 | | | | |

| | 0.55 | 2.855 | | | |
|---|---|---|---|---|---|
| (EOA)$_2$PbI$_4$ | 0.25 | 1.967 | | 1.395 | 2.41 |
| | **0.40** | 2.364 | | | |
| | 0.41 | 2.391 | | | |
| | 0.475 | 2.570 | | | |

# 5. Summary of Computational Methods

## I. Discussion of the Benchmark Data Set

The 16 2D hybrid organic-inorganic perovskites (HOIPs) for the hybrid density functional theory (DFT) benchmark were selected based on three criteria A) the availability of an absorption or emission spectrum, from which we obtained the energy of the inorganic exciton peak, and B) the availability of atomic structures without disorder for the organic atoms or atomic structure models that had been found to give a reasonable structure guess. C) Additionally, we have selected 2D HOIPs to span a variety of organic cations (see Figure S1).

The aim of this work is to predict the inorganic exciton peak energy using the Kohn-Sham band gap and the experimentally determined relationship between exciton binding energy and inorganic exciton band gap[7], as shown in Eq. (1) in the main text.

For the purpose of comparing experimental values to the Kohn-Sham band gap, we need to first determine experimental estimates for the band gap. For the 2D HOIPs $(PEA)_2PbBr_4$, $(BA)_2PbI_4$, $(EOA)_2PbI_4$, we use the experimental band gaps determined by Hansen et al. [7]. For the other 2D HOIPs, we use Eq. (1) to estimate the band gaps from the inorganic exciton peak from the absorption spectrum, and, where no absorption spectrum is available, from the emission spectrum (see Table S1).

This approach means that the estimated experimental band gap $E_g^{est}$ comes with a number of uncertainties. Eq. 1 was determined from the correlation between the exciton binding energy for 2D HOIPs and the band gap with a root-mean-square error (RMSE) of 31.5 meV. This corresponds to a variance of ~3 nm at the low energy (380 nm) of the visual spectrum and ~16 nm at the high end (750 nm).

For our benchmark, we obtain the exciton energy from experimental absorption or emission spectra (see Table S1). This will introduce additional errors into the data. First, the exciton energy needs to be extrapolated from the spectra if it is not measured more precisely in the publication, leading to a read-off error of ~0.5-1 nm. Second, the absorption and emission of the inorganic exciton peak cannot be expected to be identical due to relaxation processes. From Table S1, we can see that the difference between absorption and emission inorganic exciton peak position can be up to ~9 nm.

## II. Methods for DFT Benchmark of HSE06 functional to Model Inorganic Optical Gap

We perform all calculations within the FHI-aims all electron electronic structure code[26]. For perovskites 5 $(PEA)_2PbBr_4$, 15 $(BA)_2PbI_4$ and 16 $(EOA)_2PbI_4$, where experimental structures without disorder have been published, we performed geometry optimizations using the generalized gradient approximation (GGA) PBE functional[24] with Tkatchenko-Scheffler van-der-Waals correction [25] (TS). PBE+TS has in the past been shown to give good approximations for the atomic structure[11]. We relax the initial structure first with 'light' settings and then until forces are converged up to 0.005 eV/Å with FHI-aims's 'tight' settings. We find (Table S2) that the relaxed lattice parameters are in good agreement and within 2% of the experimental lattice parameter and angles for all but (BA)$_2$PbI$_4$, where the lattice parameter in c-direction disagrees by 2.4 %. Our deviations are in line with similar uncertainties of previous 2D HOIP structure optimizations[11,15,17].

For cross validation, we split the dataset into a training set of 12 and a test set of 4 points (one for X = Cl, one for X = Br and 2 for X = I, randomly picked from the subset). Our test set contains perovskites 3 (NMA)$_2$PbCl$_4$, 5 (PEA)$_2$PbBr$_4$, 11 (2T)$_2$PbI$_4$ and 12 (BTm)$_2$PbI$_4$.

To describe the electronic structure, we use the HSE06 functional[9] with 2$^{nd}$ variational non-self-consistent SOC[10]. For elements in the 6$^{th}$ period, the influence of spin orbit coupling (SOC) cannot be neglected when modelling the electronic structure and band gap of 2D HOIP. For Pb, the *p*-orbitals that contribute to the conduction band are split up into a $p_{1/2}$ and $p_{3/2}$ contribution due to SOC, majorly changing the electronic structure of the conduction band. We assessed in Ref.[18] the 2$^{nd}$ variational approximation non-self-consistent spin-orbit coupling approach against self-consistent SOC including $p_{1/2}$-contribution at the example of Cs$_2$AgBiCl$_6$., finding that 2$^{nd}$ variational non-self-consistent SOC underestimates the expected SOC splitting and overestimates the band gap, but reproduces the qualitative features of the band structure. This overestimation of the band gap means that the average Hartree Fock mixing parameter from our benchmark $\beta = 0.47$ is only strictly valid for HSE06+2$^{nd}$ variational non-self-consistent SOC (HSE06+SOC hereafter) as implemented FHI-aims. The linear relations given in the second half of the main text between $\beta$, the experimental band gap and the PBE+SOC band gap should however make it possible to estimate $\beta$ for other SOC approximations without repeating our benchmark.

It is an unfortunately sometimes repeated misconception that PBE without SOC can be used to get a good approximation of the experimental, inorganic band gap due to the error compensation of the self-interaction error and the neglect of SOC. In Figure S A, we show that using PBE to model the band gap (apart from errors in the electronic structure) leads to underestimating the experimental band gap by ~680 meV on average for the dataset of the 16 HOIP used in this work.

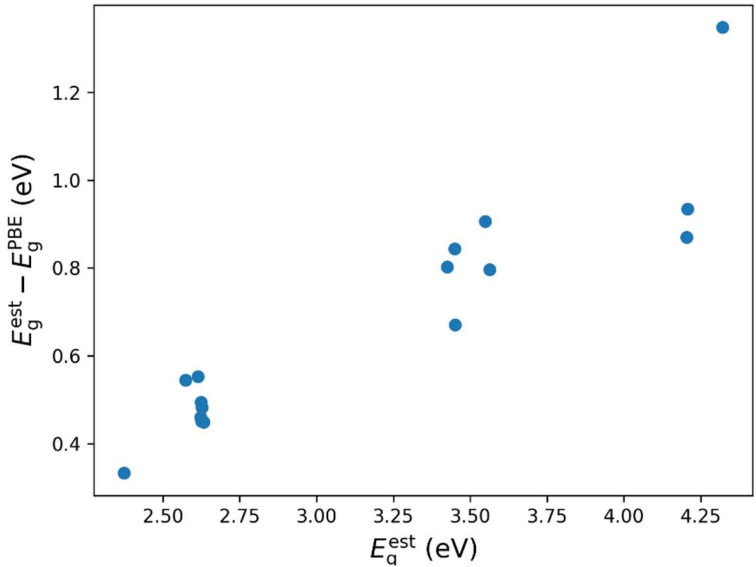

**Figure A**: The estimated, experimental inorganic band gap $E_g^{est}$ for the 16 perovskites from Table S1 as a function of the difference between $E_g^{est}$ and the inorganic band gap calculated with PBE *without spin-orbit coupling* $E_g^{PBE}$ (blue points). If $E_g^{PBE}$ were able to reproduce the inorganic band gap, the blue points would be expected to lie at 0.0 eV. This figure shows that PBE without spin-orbit coupling severely underestimates the inorganic band gap by 330 meV to 1.35 eV and on average by 680 meV. The neglect of SOC and the self-interaction error do not compensate and using PBE alone to model the inorganic band gap does not work.

We benchmark the Hartee Fock exchange parameter $\beta$ for HSE06+SOC to reproduce the extrapolated experimental band gaps of the 2D HOIPs in our benchmark set. HSE06 is defined as[9]:

$$E_{xc}^{HSE06} = \beta E_x^{HF,SR}(\omega) + (1-\beta)E_x^{PBE,SR}(\omega) + E_x^{PBE,LR}(\omega) + E_C^{PBE} \qquad (1)$$

Where $E_{xc}^{HSE06}$ is the energy from the HSE06 functional, $E_x^{HF,SR}$ is the short-range Hartree Fock exchange depending on the screening parameter $\omega$ = 0.11 bohr$^{-1}$, $E_x^{PBE,SR}$ and $E_x^{PBE,LR}$ are the short- and long-range components of the PBE exchange functional and $E_C^{PBE}$ is the PBE correlation energy. We use 'intermediate' settings within FHI-aims (we tested that 'tight' settings give the same result for the inorganic band gap for a few test cases, see Table S3).

In 2D HOIP, organic and inorganic compounds can both contribute to the band edges. We performed a Mulliken analysis[39] of the system, where we used an occupation of 0.005 per state as a threshold to assign the highest and lowest occupied states of the organic and inorganic component. We verified this approach visually against projected density of states analysis.

# 6. Figure S2: Bootstrap for the Optimized Hartree Fock Exchange Mixing Parameter $\beta$

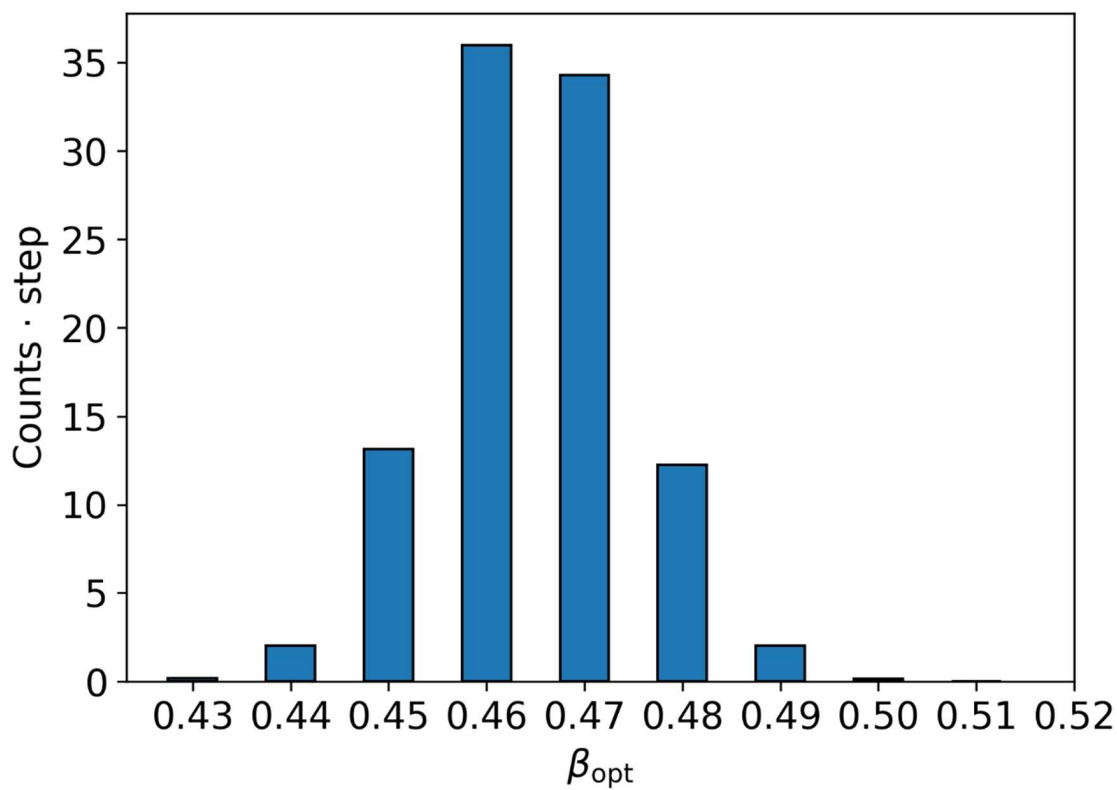

**Figure S2**: To estimate how large the variance of the optimized Hartree Fock exchange mixing parameter $\beta_{\text{opt}}$ for HSE06+SOC is, we performed a bootstrap. We took 10000 samples from the training set of 12 2D HOIPs. The step is 0.01. The resulting variance is 0.0001. Based on this bootstrap, the optimal $\beta$ for other 2D HOIPs is likely to be close to $\beta_{\text{average}} = 0.47$.

# 7. Figure S3: Level Alignment for HSE06+SOC with $\beta = 0.25$ and $\beta_{opt}$

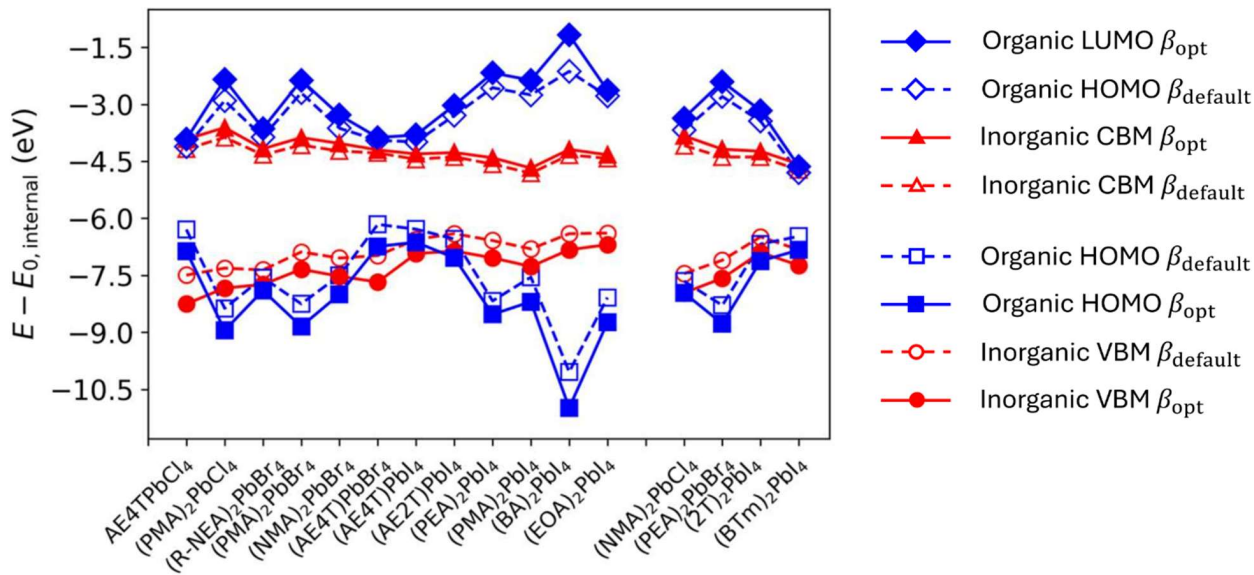

**Figure S3**: The energy of the organic (blue) and inorganic (red) frontier orbitals with respect to FHI-aims's internal zero in energy $E_{0,\text{Internal}}$ for HSE06+SOC with $\beta_{\text{default}} = 0.25$ (dashed lines with open symbols) and $\beta_{\text{opt}}$ (solid lines with filled symbols). The level alignment between the organic HOMO (blue squares) and inorganic valence band maximum (VBM, red circles) and between the organic LUMO (blue diamonds) and inorganic conduction band minimum (CBM, red triangles) remains the same between $\beta_{\text{default}}$ and $\beta_{\text{opt}}$. Note that while we have separated out the training (left) and test set (right), we are showing $\beta_{\text{opt}}$ for both training and test set here. Given that there is no change in level alignment between $\beta_{\text{default}}$ and $\beta_{\text{opt}}$, $\beta_{\text{average}}$ will maintain the level alignment, too.

# 8. Methods: Estimate of High-field Dielectric Constant of Organic and Inorganic Component from Dielectric profile

The obvious problem about obtaining the dielectric constant for the inorganic component is that organic and inorganic form one system. One approach to solve this problem is to calculate the total high-field dielectric constant and extrapolate $\varepsilon_{\infty,\text{inorg}}$[40]. A second approach is to calculate the dielectric profile[37,41]. The latter approach has in the past, e.g., been used to obtain dielectric constant for organic and inorganic contribution to estimate the inorganic exciton binding energy in dielectric models, e.g. based on Keldysh theory[42]. To obtain the dielectric profile, seven steps are necessary, which are describe below, extending the procedure outlined by Sapori et al.[37].

We use PBE with a dipole correction in FHI-aims for the underlying calculations of the dielectric profile, because our aim is to be able to obtain $\beta_{\text{opt}}$ without previously knowing $\beta_{\text{opt}}$ without having to resort to more costly HSE06 calculations. Figure S C shows that PBE gives rise to slightly higher high-field dielectric constants than HSE06 with $\beta = 0.25$. Importantly, however, the ratio between the organic and inorganic high-field dielectric constants remains the same for both functionals. We do not include SOC in our calculations, because the 2$^{\text{nd}}$ variational non-self-consistent SOC approximation implemented in FHI-aims[10] is non-self-consistent, i.e., the orbitals and electron density are not self-consistently updated under the SOC approximation.

The steps to obtain $\varepsilon_{\infty,\text{inorg}}$ and $\varepsilon_{\infty,\text{org}}$ are:

1. The unit cell for the calculations needs to be determined. The bulk 2D HOIP needs to be converted into a slab model where the surface is perpendicular to the organic and inorganic layers of the 2D HOIP (assumed here to be in z-direction). The first challenge with this step is that the angle between the lattice vector in z-direction can differ from 90°. If this is the case, the unit cell needs to be rotated such that this tilt is reflected by the z-lattice vector only (i.e., no z-components for the x- and y- lattice vector). After a unit cell with the desired number of layers has been built, the x and y components of the z-lattice vector have to be deleted to allow for insertion of the vacuum layer. The second problem is how many layers of organic and inorganic component should be included into the slab, which can be determined by convergence test. We have used unit cells of the configuration of organic-inorganic-organic-inorganic-organic layers. For (EOA)$_2$PbI$_4$, we found that this configuration gave converged dielectric constants for the inner inorganic and organic layers (see Figure S D).
2. The total electron density is calculated using PBE and 'tight' settings, with and without application of an external field $E_{\text{ext}} = 0.1$ eV using a step of 0.1 Å in z-direction and a step of 0.05 Å in x and y direction to print out cube files. The electron density is then averaged perpendicular to the slab plane (i.e., along the vacuum axis, labeled z here), which results in the electron density profile:

$$\rho(z) = \frac{1}{V(z)} \sum_x^{n_x} \sum_y^{n_y} \rho(x,y,z) dV \qquad (2)$$

Here $n_x, n_y$ are the number of voxels in $x$ and $y$ dimension, $dV$ is the volume of a voxel in the cube file with the side lengths $dx, dy, dz$. $V(z) = \sum_x^{n_x} \sum_y^{n_y} dV$.

3. The induced electron density $\delta p_{\text{ind}}(z)$ is calculated by subtracting the electron density profile without field from the electron density with applied external field.

$$\delta p_{\text{ind}}(z) = \rho(z)|_{E_{\text{ext}}} - \rho(z)|_0 \tag{3}$$

4. The induced electron density then needs to be corrected for unphysical inter-slab interaction. FHI-aims *allows us to correct for the induced dipole moment already at the step of the PBE calculations using the keyword use_dipole_correction*, making it unnecessary to apply this correction to induced electron density. To obtain the correction for the usecase without use_dipole_correction, the total induced dipole $m_z$ can be calculated by integrating over $z$ according to:

$$m_z = \int_{z_{\text{min}}}^{z_{\text{maz}}} p_{\text{ind}}(z) \, dz \tag{4}$$

for three unit cells with different vacuum layer thickness L. Then, the inverse induced dipole moment is plotted as a function of the inverse vacuum layer thickness, $\frac{1}{m_z} = \frac{a}{L} + m_{z,\text{inf}}$. The slope corresponds to $a = 1/(E_{\text{ext}} \varepsilon_0)$ [43] and the y-intercept corresponds to the induced dipole moment at infinite vacuum level ($m_{z,\text{inf}}$). The induced electron density can then be corrected by multiplying $\delta p_{\text{ind}}(z)$ using a factor of $m_{z,\text{inf}}/m_z$. This correction is only necessary if the DFT calculations cannot already be corrected for the unphysical slab-interaction in z-direction.

5. The induced polarization can then be calculated as:

$$\frac{dp_{\text{ind}}(z)}{dz} = -\delta \rho_{\text{ind}}(z) \tag{5}$$

6. Then, dielectric profile can be calculated using the vacuum permittivity $\varepsilon_0$ ($\varepsilon_0 = 0.005526349406$ e$^2$ eV$^{-1}$Å$^{-1}$) according to:

$$\varepsilon_\infty(z) = \frac{\varepsilon_0 E_{\text{ext}}}{\varepsilon_0 E_{\text{ext}} - p_{\text{ind}}(z)} \tag{6}$$

7. Integrating the inverse dielectric profile can be used to calculate the inverse, effective, static dielectric constant[41,44].

$$\frac{1}{\varepsilon_{\infty,\text{eff}}} = \frac{1}{z_2 - z_1} \sum_{i=z_1}^{z_2} \frac{1}{\varepsilon_\infty(z)} dz \tag{7}$$

Note that the dielectric constant depends on the integration borders $z_2$ and $z_1$. As the dielectric profile between organic and inorganic is continuous and because the organic linker often overlaps with the axial halide positions of the inorganic in $z$-direction, the choice of where to separate the organic and inorganic component is not straightforward. We use the position of the axial halide atoms extended by $t_1$ Å into the organic component as separation between the organic and inorganic component. We find that it does not influence the dielectric constant of organic and inorganic significantly (see Tables S4 and S5 below). A better approach could be to calculate the dielectric profile of the organic component without presence of the inorganic component (removing one H-atom from each linker group to obtain charge neutrality) and subtracting the organic dielectric profile from the organic-inorganic profile.

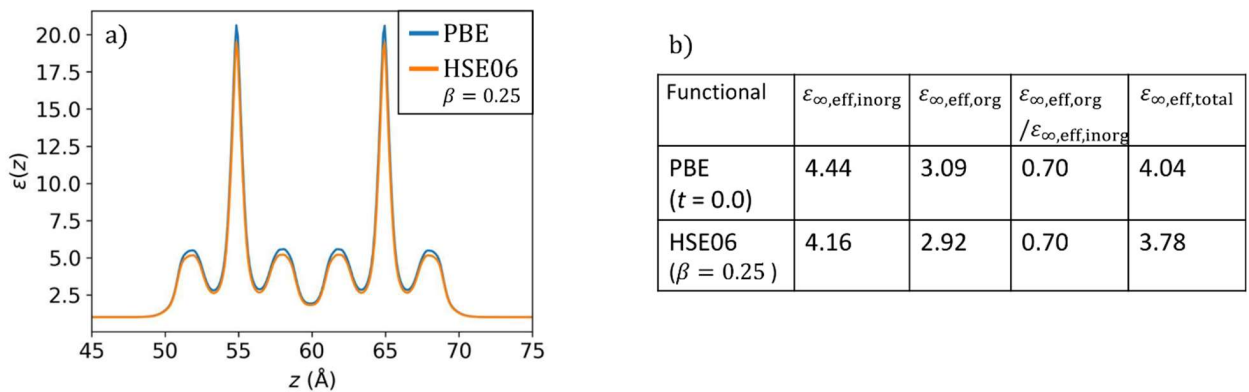

**Figure S C**: We investigate the dependence of the dielectric constant on different functionals at the example of (EOA)$_2$PbI$_4$. PBE gives slightly higher values for the high-field dielectric constant compared to HSE06 with $\beta = 0.25$. For both calculations, $t_1 = 0.0$ Å is used. The ratio between organic and inorganic high-field dielectric constant remains the same.

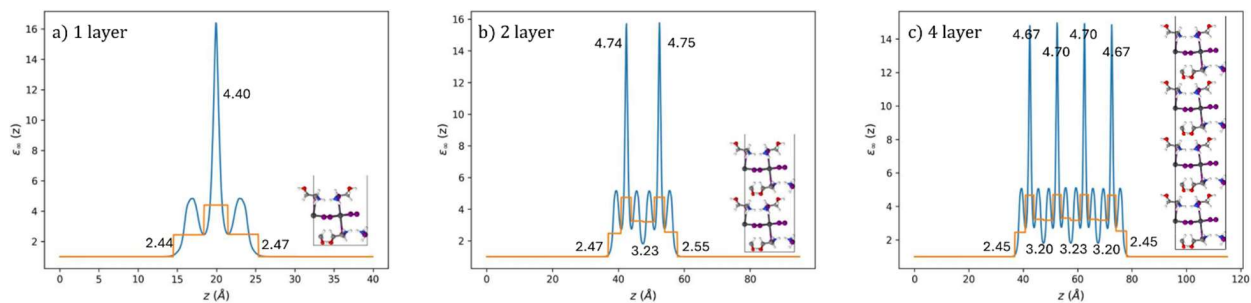

**Figure S D**: We investigated the dielectric constant for different layers of (EOA)$_2$PbI$_4$. We tested different slab sizes for (EOA)$_2$PbI$_4$ and define a layer as the configuration (EOA)-PbI$_4$-(EOA) (pictures of unit cells are shown as insets in each panel, modeled with jmol[45]). We investigated a) 1 layer with a ratio of the organic and inorganic constant of the inner layers of $\varepsilon_{\infty,\text{eff,org}}/\varepsilon_{\infty,\text{eff,inorg}} = 0.56$, b) 2 layers with $\varepsilon_{\infty,\text{eff,org}}/\varepsilon_{\infty,\text{eff,inorg}} = 0.68$, c) 4 layers with $\varepsilon_{\infty,\text{eff,org}}/\varepsilon_{\infty,\text{eff,inorg}} = 0.68$ and found that the dielectric constants for organic and inorganic beyond the outer organic layers and in particular their ratio $\varepsilon_{\infty,\text{eff,org}}/\varepsilon_{\infty,\text{eff,inorg}}$ are converged between 2 and 4 layers, meaning that 2 layers should be sufficient for our investigations. Note that the calculations were done by correction for the inter slab dipole according to the procedure described in step 4, so without the use_dipole_correction keyword and while using 'intermediate' settings, leading to sightly different values in the dielectric constants compared to Tab. S4 and S5.

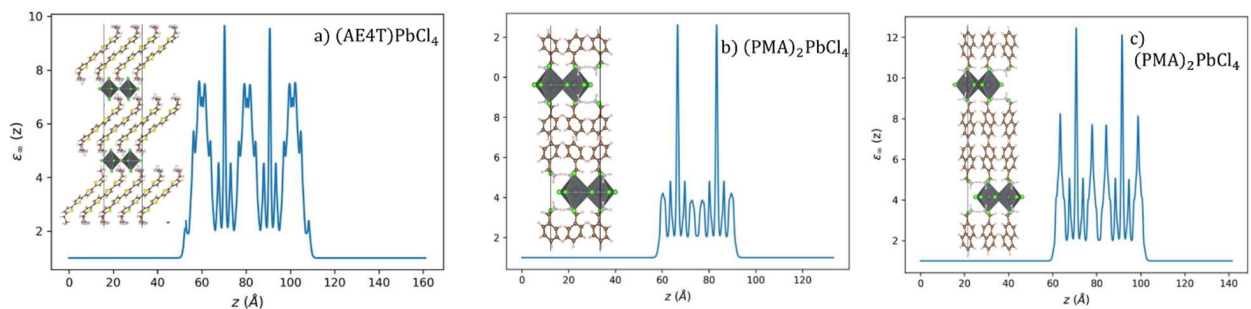

**Figure S E**: Dielectric profiles for all the Cl-based 2D HOIPs with picture of the slabs as an inset a) (AE4T)PbCl$_4$ b) (PMA)$_2$PbCl$_4$ c) (NMA)$_2$PbCl$_4$. Insets show unit cells generated with VESTA[34].

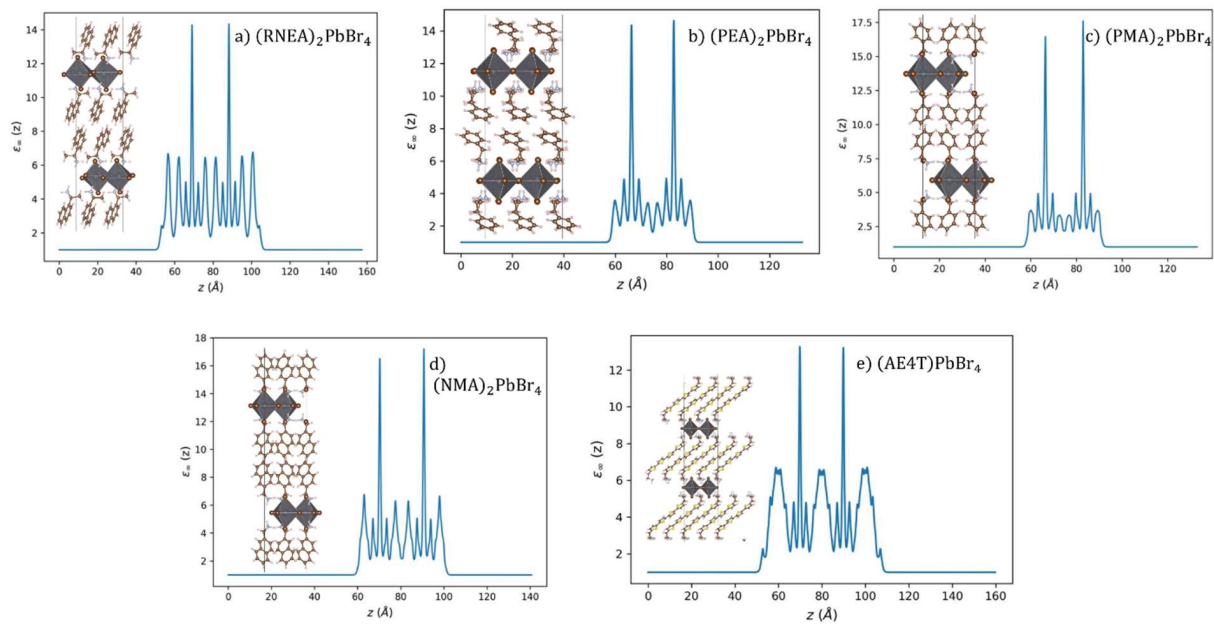

**Figure S F**: Dielectric profiles for all the Br-based 2D HOIPs with picture of the slabs as an inset a) (R-NEA)$_2$PbBr$_4$ b) (PEA)$_2$PbBr$_4$, c) (PMA)$_2$PbBr$_4$, d) (NMA)$_2$PbBr$_4$, e) (AE4T)PbBr$_4$. Insets show unit cells generated with VESTA[34].

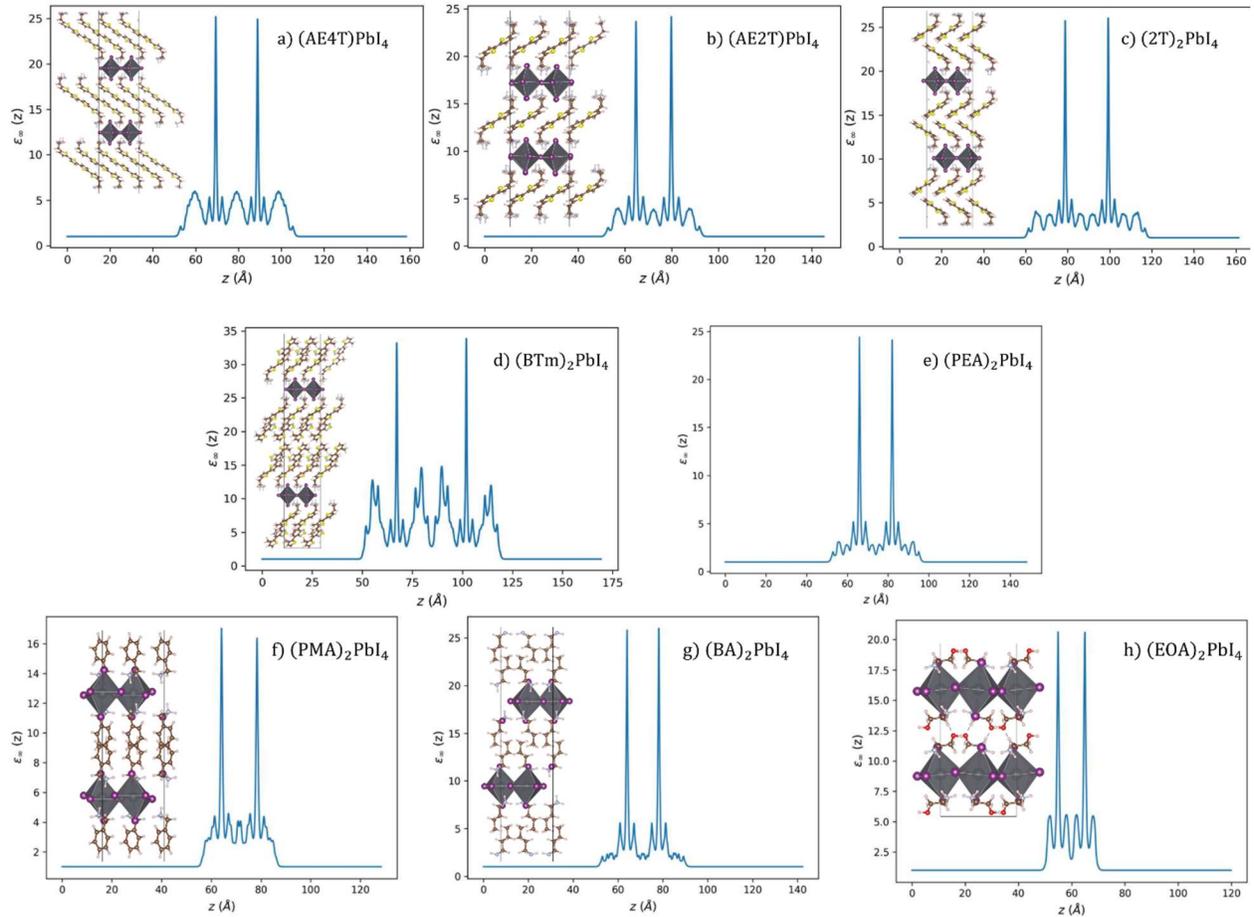

**Figure S G**: Dielectric profiles for all the I-based 2D HOIPs with picture of the slabs as an inset e) (AE4T)PbI$_4$, b) (AE2T)PbI$_4$. c) (2T)$_2$PbI$_4$ d) (BTm)$_2$PbI$_4$, e) (PEA)$_2$PbI$_4$, f) (PMA)$_2$PbI$_4$, g) (BA)$_2$PbBr$_4$, h) (EOA)$_2$PbI$_4$. Insets show unit cells generated with VESTA[34].

The dielectric constants we obtained for the organic and inorganic component are shown for different cut-offs between organic and inorganic component in Tables S4 and S5. The difference between the dielectric values for cut-offs $t_1 = 0.0$ Å and for $t_1 = 0.1$ Å is not very big, particularly the organic-inorganic dielectric mismatch $\varepsilon_{\infty,\text{eff,org}}/\varepsilon_{\infty,\text{eff,inorg}}$ do not change significantly. As mentioned above and shown in Figure S C, at the example of (EOA)$_2$PbI$_4$, PBE is expected to give slightly higher values for the high-field dielectric constant than HSE06. Because the difference between the functionals is small and affects $\varepsilon_{\infty,\text{eff,org}}$ and $\varepsilon_{\infty,\text{eff,inorg}}$ in the same manner as can be seen by the constant ratio $\varepsilon_{\infty,\text{eff,org}}/\varepsilon_{\infty,\text{eff,inorg}}$, the choice of the functional should not significantly influence our results, apart from one exception: For (BTm)$_2$PbI$_4$ , $\varepsilon_{\infty,\text{eff,org}}$ and $\varepsilon_{\infty,\text{eff,inorg}}$ are much larger than for the other 2D HOIPs. The reason for this is likely that due to the self-interaction error, PBE underestimates the (organic) overall band gap of the material and perceive the material as metallic when $E_{\text{ext}} = 0.1$ eV is applied, leading to a wrong charge-profile for $E_{\text{ext}} = 0.1$ eV. This assumptions is born out by the fact that $\varepsilon_{\infty,\text{eff,org}}/\varepsilon_{\infty,\text{eff,inorg}}$ falls into the same and expected range as for the other 2D HOIPs: a constant off-set resulting from an error in one of the charge profiles cancels out. Clearly, when the self-interaction error can be expected to affect results,

HSE06 needs to be chosen over PBE. For the purpose of this work, we have excluded (BTm)$_2$PbI$_4$ from the evaluation.

As remarked in the main text, we find that the dielectric mismatch between organic and inorganic component $\varepsilon_{\infty,\text{eff,org}}/\varepsilon_{\infty,\text{eff,inorg}}$ does not appear to be very large for any of the 2D HOIPs in agreement with similar recent observations of Hansen[7], going against the previous common assumption of a large organic/inorganic dielectric mismatch. In fact, for (PMA)$_2$PbCl$_4$, (NMA)$_2$PbCl$_4$, (R-NEA)$_2$PbBr$_4$, (NMA)$_2$PbBr$_4$, (AE4T)PbBr$_4$, (AE4T)PbI$_4$, the dielectric constants are almost identical.

Additionally, we observe that the dielectric constant for the same organic cation (AE4T$^{2+}$, PMA$^+$, PEA$^+$, NMA$^+$) changes slightly depending on the inorganic cation. This change is not surprising, because the bond lengths Pb-X increase from X = Cl to I, meaning that the organic cations need to fill out more space from X = Cl to I, affecting their orientation. This difference in orientation and room-filling behavior can be seen when comparing the atomic structure from the insets in Figures S E-G.

In previous investigations[7], the high-field dielectric constant was obtained from independent quasiparticle approximation for (PEA)$_2$PbBr$_4$ (2.74), (PEA)$_2$PbI$_4$ (3.29), (BA)$_2$PbI$_4$ (3.30), (EOA)$_2$PbI$_4$ (3.94). These values agree reasonably well with the high-field dielectric constant we obtain here for $t_1 = 0.0$ Å of (PEA)$_2$PbBr$_4$ (3.05), (PEA)$_2$PbI$_4$ (3.29), (BA)$_2$PbI$_4$ (3.30) and (EOA)$_2$PbI$_4$ (4.04). The agreement is on first glance not quite as satisfying when looking at the estimated values for the organic and inorganic high-field dielectric constant from literature. EOA$^+$ and PEA$^+$ in 2D HOIPs were determined to be 2.45 and 2.40 [7], respectively, smaller than we find in our investigations for $t_1 = 0.0$ Å (EOA: 3.09, PEA: 2.6). Similarly, The high-field dielectric constant for the inorganic components [PbCl$_4^{2-}$], [PbBr$_4^{2-}$] and [PbI$_4^{2-}$] were determined as 2.79, 3.45 and 4.68[7]. Compared to our average high-field dielectric constants $\varepsilon_{\infty,\text{eff,inorg}}$ of 3.17, 3.56 and 4.14, we obtain larger values for X = Cl, good agreement for X = Br and lower values for X = I.

Given the approximations in our approach (using PBE and cutting the dielectric profile sharply to attribute the organic and inorganic dielectric constant) and given that the experimentally reported values for $\varepsilon_{\infty,\text{eff,org}}, \varepsilon_{\infty,\text{eff,inorg}}$ were fitted from the total dielectric constant using the relation between the dielectric constants and the length of the organic and inorganic in *z*-distance of Hong et al.[7,40], the agreement is however very good, as the same trends are obtained. Our agreement is similar as observed previously for the same approach to extract the dielectric constants from the dielectric profile: For (MA)PbBr$_3$, Even et al.[37] obtained a high-field dielectric constant of (MA)PbBr$_3$ 4.2 and for (MA)PbI$_3$ of 5.1 compared to the experimental value of 3.55 and 4.53, respectively[7,46].

Figure S5 shows $\varepsilon_{\infty,\text{total}}$ and $\varepsilon_{\infty,\text{eff,org}}/\varepsilon_{\infty,\text{eff,inorg}}$ as a function of the inverse, optimize $\beta_{\text{opt}}$. It is clear that there is no correlation between $1/\beta_{\text{opt}}$ and $\varepsilon_{\infty,\text{total}}$ or $\varepsilon_{\infty,\text{eff,org}}/\varepsilon_{\infty,\text{eff,inorg}}$. It is unlikely that this lack in correlation is a result of how we set the integration borders: To obtain a clearer correlation, the dielectric constant of the organic would either need to be smaller or the inorganic dielectric constant larger. Pushing the integration border further into the organic component does not seem sensible, because the organic tether group is for many of the 2D HOIPs already reaching into the inorganic component. In fact, for $t_1 = 0.1$ Å, the dielectric constant of the inorganic has become smaller. Taking narrower integration borders would mean excluding the axial inorganic halide atom. Traore et al.[47] showed that the 2D HOIP dielectric profile can be seen as a sum of the contributions of the constituting components. Hence, an approach to estimate the dielectric contribution of the inorganic could be to calculate the dielectric contribution of the organic separately (by removing the inorganic component and one H-atom per organic tether group to obtain charge neutrality) and then subtract the organic dielectric profile from the total dielectric profile. However, this approach would again require choosing integration borders for Eq. (7) and would add a significant amount of complexity and computational effort. It seems therefore clear that no correlation between $\beta_{\text{opt}}$ and the 2D HOIPs dielectric constants exists.

# 9. Figure S4: Correlation of total dielectric constant and dielectric mismatch with 1/β

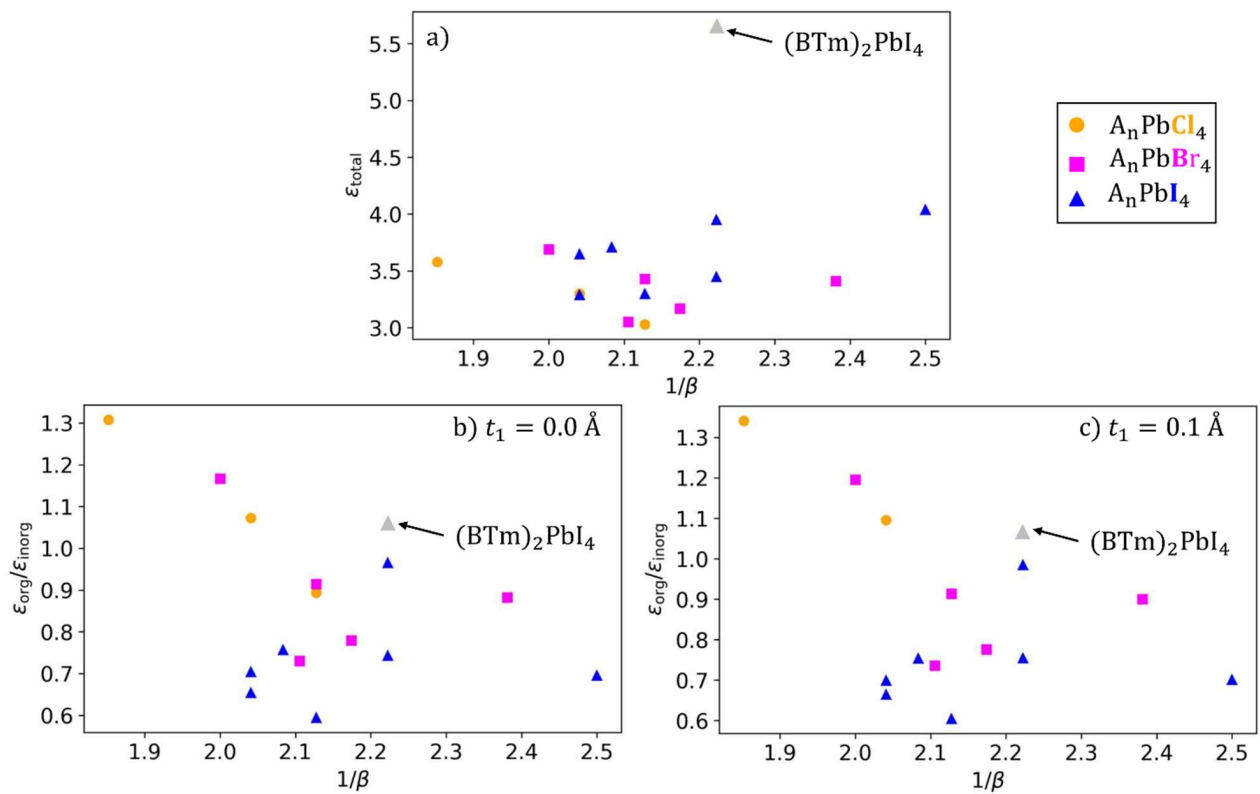

**Figure S4**: a) Total high-field dielectric constant $\varepsilon_{\infty,\text{total}}$ vs $1/\beta_{\text{opt}}$; b) Dielectric mismatch $\varepsilon_{\infty,\text{eff,org}}/\varepsilon_{\infty,\text{eff,inorg}}$ vs $1/\beta_{\text{opt}}$ for different cut-offs in organic and inorganic dielectric profile $t_1 = 0.0$ Å and b) $t_1 = 0.1$ Å. There is no clear correlation between the total high-field dielectric constant or the organic-inorganic dielectric mismatch and the optimized $\beta_{\text{opt}}$. For (BTm)$_2$PbI$_4$, under $E_{\text{ext}} = 0.1$ eV, calculations with PBE wrongly show no band gap (a consequence of the self-interaction error), leading to the too high values for the different $\varepsilon_{\infty,\text{eff}}$ shown here. We have excluded (BTm)$_2$PbI$_4$ from our evaluations, but show the values in this figure as a grey triangle for completeness sakes.

# 10. Table S4: High-field Effective Dielectric Constants from Dielectric Profile for $t_1$ = 0.0 Å

**Table S4**: Dielectric constants extracted from the PBE dielectric profile for organic and inorganic and organic/inorganic mismatch $\varepsilon_{\infty,\text{eff,org}}/\varepsilon_{\infty,\text{eff,inorg}}$. The dielectric profiles were separated into organic and inorganic component at the position of the axial halide atoms. For (BTm)$_2$PbI$_4$, under $E_{\text{ext}} = 0.1$ eV, calculations with PBE wrongly show no band gap (a consequence of the self-interaction error), likely leading to the too high values for the different $\varepsilon_{\infty,\text{eff}}$ shown below. We have hence excluded (BTm)$_2$PbI$_4$ from our evaluations.

| Perovskite | $\varepsilon_{\infty,\text{eff,inorg}}$ | $\varepsilon_{\infty,\text{eff,org}}$ | $\varepsilon_{\infty,\text{eff,org}}/\varepsilon_{\infty,\text{eff,inorg}}$ | $\varepsilon_{\infty,\text{eff,total}}$ |
|---|---|---|---|---|
| (AE4T)PbCl$_4$ | 3.12 | 4.08 | 1.31 | 3.58 |
| (PMA)$_2$PbCl$_4$ | 3.20 | 2.86 | 0.89 | 3.03 |
| (NMA)$_2$PbCl$_4$ | 3.18 | 3.41 | 1.07 | 3.30 |
| (R-NEA)$_2$PbBr$_4$ | 3.65 | 3.22 | 0.88 | 3.41 |
| (PEA)$_2$PbBr$_4$ | 3.56 | 2.6 | 0.73 | 3.05 |
| (PMA)$_2$PbBr$_4$ | 3.58 | 2.79 | 0.78 | 3.17 |
| (NMA)$_2$PbBr$_4$ | 3.60 | 3.29 | 0.91 | 3.43 |
| (AE4T)PbBr$_4$ | 3.42 | 3.99 | 1.17 | 3.69 |
| (AE4T)PbI$_4$ | 4.01 | 3.87 | 0.97 | 3.95 |
| (AE2T)PbI$_4$ | 4.22 | 2.97 | 0.70 | 3.65 |
| (2T)$_2$PbI$_4$ | 4.05 | 3.01 | 0.74 | 3.45 |
| (BTm)$_2$PbI$_4$ | 5.44 | 5.76 | 1.06 | 5.65 |
| (PEA)$_2$PbI$_4$ | 4.02 | 2.63 | 0.65 | 3.29 |
| (PMA)$_2$PbI$_4$ | 4.16 | 3.15 | 0.76 | 3.71 |
| (BA)$_2$PbI$_4$ | 4.07 | 2.42 | 0.59 | 3.30 |
| (EOA)$_2$PbI$_4$ | 4.44 | 3.09 | 0.70 | 4.04 |

# 11. Table S5: High-field Effective Dielectric Constants from Dielectric Profile for $t_1 = 0.1$ Å

**Table S5**: Dielectric constants extracted from the dielectric profile for organic and inorganic and organic/inorganic mismatch $\varepsilon_{\infty,\text{eff},\text{org}}/\varepsilon_{\infty,\text{eff},\text{inorg}}$. The dielectric profiles were separated into organic and inorganic component $t_1 = 0.1$ Å away from the position of the axial halide atoms into the organic layer. For (BTm)$_2$PbI$_4$, under $E_{\text{ext}} = 0.1$ eV, calculations with PBE wrongly show no band gap (a consequence of the self-interaction error), likely leading to the too high values for the different $\varepsilon_{\infty,\text{eff}}$ shown below. We have hence excluded (BTm)$_2$PbI$_4$ from our evaluations..

| Perovskite | $\varepsilon_{\infty,\text{eff},\text{inorg}}$ | $\varepsilon_{\infty,\text{eff},\text{org}}$ | $\varepsilon_{\infty,\text{eff},\text{org}}/\varepsilon_{\infty,\text{eff},\text{inorg}}$ | $\varepsilon_{\infty,\text{eff},\text{total}}$ |
|---|---|---|---|---|
| (AE4T)PbCl$_4$ | 3.08 | 4.13 | 1.34 | 3.56 |
| (PMA)$_2$PbCl$_4$ | 3.15 | 2.88 | 0.91 | 3.02 |
| (NMA)$_2$PbCl$_4$ | 3.13 | 3.43 | 1.10 | 3.29 |
| (R-NEA)$_2$PbBr$_4$ | 3.60 | 3.24 | 0.90 | 3.40 |
| (PEA)$_2$PbBr$_4$ | 3.52 | 2.59 | 0.74 | 3.05 |
| (PMA)$_2$PbBr$_4$ | 3.58 | 2.78 | 0.78 | 3.17 |
| (NMA)$_2$PbBr$_4$ | 3.60 | 3.29 | 0.91 | 3.42 |
| (AE4T)PbBr$_4$ | 3.37 | 4.03 | 1.20 | 3.67 |
| (AE4T)PbI$_4$ | 3.96 | 3.9 | 0.98 | 3.93 |
| (AE2T)PbI$_4$ | 4.22 | 2.95 | 0.70 | 3.65 |
| (2T)$_2$PbI$_4$ | 4.00 | 3.02 | 0.76 | 3.44 |
| (BTm)$_2$PbI$_4$ | 5.42 | 5.77 | 1.06 | 5.65 |
| (PEA)$_2$PbI$_4$ | 3.96 | 2.63 | 0.66 | 3.28 |
| (PMA)$_2$PbI$_4$ | 4.15 | 3.13 | 0.75 | 3.71 |
| (BA)$_2$PbI$_4$ | 4.00 | 2.42 | 0.61 | 3.25 |
| (EOA)$_2$PbI$_4$ | 4.39 | 3.08 | 0.70 | 4.03 |